# Transition behavior of the seizure dynamics modulated by the astrocyte inositol triphosphate noise

Jiajia Li[1, 2, a)], Peihua Feng[3], Liang Zhao[1], Junying Chen[1], Mengmeng Du[4], Jian Song[2, b)] and Ying Wu[3]

[1)] College of Information and Control Engineering, Xi'an University of Architecture and Technology, Shaanxi, Xi'an, 710055, China;
[2)] Department of Neurosurgery, Wuhan General Hospital of PLA, Wuhan 430070, China
[3)] State Key Laboratory for Strength and Vibration of Mechanical Structures, National Demonstration Center for Experimental Mechanics Education, School of Aerospace Engineering, Xi'an Jiaotong University, Xi'an 710049, China
[4)] School of Mathematics and Data Science, Shaanxi University of Science & Technology, Xi'an 710021, China

**ABSTRACT**

Epilepsy is a neurological disorder with recurrent seizures, which convey complex dynamical characteristics including chaos and randomness. Until now, the underlying mechanism has not been fully elucidated, especially the bistable property beneath the epileptic random induction phenomena in certain conditions. Inspired by the recent finding that astrocyte GTPase-activating protein (G-protein)-coupled receptors could be involved in stochastic epileptic seizures, we proposed a neuron-astrocyte network model, incorporating the noise of the astrocytic second messenger, inositol triphosphate (IP3) which is modulated by the G-protein-coupled receptor activation. Based on this model, we have statistically analysed the transitions of epileptic seizures by performing repeatable simulation trials. Our simulation results show that the increase of the IP3 noise intensity induces the depolarization-block epileptic seizures together with an increase in neuronal firing frequency, consistent with corresponding experiments. Meanwhile, the bistable states of the seizure dynamics were present under certain noise intensities, during which the neuronal firing pattern switches between regular sparse spiking and epileptic seizure states. This random presence of epileptic seizures is absent when the noise intensity continues to increase, accompanying with an increase in the epileptic depolarization block duration. The simulation results also shed light on the fact that calcium signals in astrocytes play significant roles in the pattern formations of the epileptic seizure. Our results provide a potential pathway for understanding the epileptic randomness in certain conditions.

---

[a)] lijiajia_dynamics@xauat.edu.cn
[b)] docsongjian@gmail.com

**The paper considers the fast-slow dynamical models of neuron- astrocyte coupled network. The model can well explain the representative case of bursting related to seizure(SZ) and depolarization block (DB). In particular, such a pattern occurred in a random pathway of astrocyte inositol triphosphate noise, and the bistable states at the critical parameters were given to explain the coexistence phenomena between spiking and SZs in certain conditions. An approximate comparasion was also performed to adapt the dynamical frequency domain of the slow subsystem of astrocyte to the experiments.**

I. INTRODUCTION

Epilepsy is a worldwide neurological disorder in which seizures occur randomly and generally demonstrate periods of unusual behaviour, sensations, and sometimes loss of awareness. Epileptic seizures are transient paroxysms of excessive discharge of neurons in brains of epileptic patients, which are generally characterized by various firing patterns ranging from the slow waves, the tonic-clonic waves to the depolarization blocks etc.[1] Over 70 million people have suffered from these different types of epileptic seizures worldwide[1], a high percent of which is difficult to be controlled[2,3]. One of the most important reasons for this is the lack of comprehensive knowledge regarding the mechanisms of epilepsy induction, including randomness during seizures in certain conditions[4,5]. Therefore, unveiling the potential mechanisms of epileptic seizure randomness is critical for obtaining a better understanding of both epileptic generation and termination. In previous methodological studies, some experts investigated the underlying dynamic characteristics based on epileptic electroencephalogram (EEG) data. For instance, Baud used the regular temporal structure of an epileptic EEG dataset to conclude against the concept that seizures are completely random events[6]. Karoly et al. then utilized a statistical method to detect significant interictal spike information among random epileptic seizure EEGs[7]. However, these EEG-based methodological studies lack the capability to explain the underpinning of epileptic randomness in view of micro biological neuronal networks.

To elucidate the dynamic mechanisms of brain activities in view of biological neuronal networks, modelling studies based on the Hodgkin-Huxley network and related mathematical frameworks have been extensively used to explain epileptic mechanisms. For example, Burn et al. constructed a complex neuronal network model with rich-hub network structures to prove the significant role of network-rich hubs in the generation of synchronous epileptic seizures[8]. Kramer utilized mean-field neural models to explain critical bifurcation phenomena during long-lasting epileptic seizures[9]. Lytton et al. also developed a computer simulation frame to successfully retrieve the dynamic transitions between epileptic tonic and clonic phases[10]. In addition, astrocytes, which have long been disregarded, have been found in recent decades to be critical for epileptic seizures and their stochastic characteristics[11–14]. The neuron-astrocyte interaction loop, termed dressed neuron, emerges to maintain highly efficient computation capacity in unfolding the mechanism of brain diseases[15]. Recent studies have developed models to study the modulatory dynamics of astrocytes in epileptic seizures in different fields. For example, Witthoft and his colleagues established models of potassium disuse in gap junction channels to propose that the astrocyte gap junction block contributes to epileptic seizures due to the degradation of potassium buffering function (16). Similarly, Ullah and other teams demonstrated that astrocytes also buffer inositol triphosphate (IP3) and glucose, which contribute to epileptic seizures[17, 18]. In view of the astrocyte glutamate modulation effect, Nadkarni and his colleagues performed a bifurcation analysis of CA3 neuron-astrocyte local systems and proved that astrocyte glutamate plays significant roles in decreasing the epilepsy generation threshold[19, 20]. Additionally, our previous modelling study proved the positive role of abnormalities in astrocyte glutamate degradation in epileptic generation[21], which was confirmed in an *in vitro* experiment[22].

Because of the high correlations between astrocytes and neuronal functions, unveiling the randomness mechanism of epileptic seizures could implicate astrocytes. As recent experiments reported, there were diverse types of G-proteins, such as Gq and Gi/o types[23]. G-protein-coupled receptors have a notable effect on astrocyte calcium ($Ca^{2+}$) signals and indirectly affect the feedback modulation of astrocytes on neurons. Wherein, the increase of IP3 concentration was reported to be the intermediate consequence for the activation of G-protein-coupled receptors and the astrocyte $Ca^{2+}$ signal, which was found to be affected by the astrocyte extracellular circumstances of mechanical, thermal and volume changing factors[24–26]. These all contribute to the noisy condition of the IP3 changing process when the neuronal glutamate binds the G-protein couple receptors of astrocytes. Therefore, there is high potential for the noisy IP3 changes induced by the astrocyte G-protein heterogeneity and other circumstance factors within the population to modulate the stochastic dynamic activities of neurons and astrocytes, which can partly explain the randomness of epileptic seizures in certain condition. To investigate the underlying mechanism by which IP3 changing noise modulates the epileptic seizures of neurons, a neuron-astrocyte network model was first developed by considering the all-to-all random connection of neurons and the IP3 changing noise model of astrocytes. Meanwhile, the entropy theory was first used to describe the evolution of epileptic seizures.

The simulation results of this study were performed as follows. By simulating the growing conditions of IP3 changing noise, we first discussed the state transitions of epileptic seizures versus the increase in the noise intensity, for each parameter condition of which, the simulations were performed 60 times to test the statistical and random property of epileptic seizures. Additionally, we studied the underlying mechanisms of the long-lasting dynamics of noise-induced epileptic seizures by analysing the correlation between neuronal firing dynamics in the amplitude and frequency domains, astrocyte long-lasting signals (intracellular calcium signals in astrocytes), and seizure duration enlargements as a consequence of the increase in noise intensity. Finally, the reasonability and the parameter values of our simulated results have been discussed with related experiments and previous theoretical studies in the discussion section.

## II. MODELS

In the last two decades, the neural field model has been developed from the globally coupled microneural oscillation model[27] to describe the macro-sclae EEG signals of the complex potential activities of brain normal cognition[28, 29] and disordered brain diseases, including epilepsy[30–34]. Epileptic seizures in multiple-unit potentials or single EEG channel potentials have been extensively studied by the mean-field neuron model. However, because of its limitation in only describing fast brain activities in milliseconds, it lacks the capacity to describe the long-lasting and random dynamics of epileptic seizures in seconds, which can be redeemed by incorporating the slow system of astrocytes; therefore, we introduced astrocytes into the model based on the basic principle of neuron-astrocyte couple systems on a micro spatial scale with chain-shaped network of *N* neurons and astrocyte respectively. Finally, the fast-slow neuron-astrocyte coupled network scheme was established and is depicted in Fig. 1.

As shown in the network in Fig. 1, on the biological basis of the neuron-astrocyte communication circuits in experiments[35, 36] and other modelling studies of neuron-astrocyte interaction circuits[37, 38], we determined that the synaptic puffs of glutamate from the other neurons in the network activate not only dendrites of the target neuron but also neighbouring astrocytes, leading to astrocytic $Ca^{2+}$-dependent feedback to dendrites. Consequently, the target neuron receives not only direct synaptic puffs but also transformed synaptic puffs from astrocytes, which results in a very high efficiency of synaptic information transmission observed in in vivo cortical studies through glutamate and GABA pathways respectively[39, 40]. Therefore, it is more cogent to use the network model to investigate epileptic seizures in contrast with seizures.

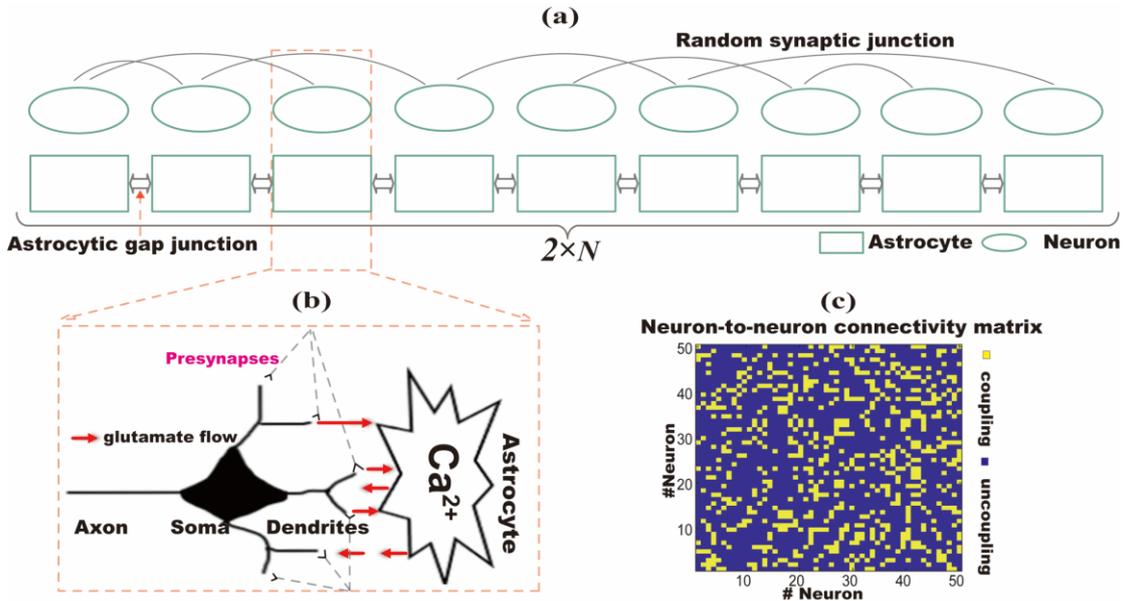

FIG. 1. The fast-slow neuron-astrocyte coupled network scheme. (a) At the top of the network, the model includes *N* pairs of neuron-astrocyte circuits. Neurons are coupled with chemical synaptic connections, and astrocytes are coupled by nonlinear gap junction channels. For the astrocyte connection channel, it allows a nonlinear diffusion of IP3 motivated by the IP3 concentration gradient among the connected astrocytes. Because the astrocyte network is chain-shaped, each astrocyte connected with the former and the latter astrocyte in the network (See the hollow double arrows) considering the limited astrocyte synaptic territory; (b) The bottom left panel shows

inhibition of the neuron-astrocyte circuit in the local network, where the presynapses from the other neurons in the network activate not only the dendrites of the target neuron but also the neighbouring astrocyte through partial glutamate (See the glutamate flow), leading to astrocytic $Ca^{2+}$-dependent feedback to the dendrites with the astrocyte-released glutamate; (c) The bottom right panel of the connection map represents the connection state across all neurons. The synaptic connection probability was chosen as 0.22, which is close to the brain cortex parameter domain.

In the network, to describe the correlation between astrocytes and neurons more reasonably, we utilized the biophysical Li-Rinzel model[41] to simulate $Ca^{2+}$ activity in each astrocyte:

$$\frac{d[Ca^{2+}]_i}{dt} = c_1 v_1 p_\infty^3 n_\infty^3 q^3 ([Ca^{2+}]_{ER} - [Ca^{2+}]_i) + c_1 v_2 ([Ca^{2+}]_{ER} - [Ca^{2+}]_i) - \frac{v_3 [Ca^{2+}]_i^2}{[Ca^{2+}]_i^2 + k_3^2} - J_{leak} \qquad (1)$$

$$\frac{dq_i}{dt} = \alpha_q (1 - q_i) + \beta_q q_i + \xi_i(t), \qquad (2)$$

$$p_\infty = \frac{[IP_3]_i}{[IP_3]_i + d_1}, \quad n_\infty = \frac{[Ca^{2+}]_i}{[Ca^{2+}]_i + d_5}; \qquad (3)$$

$$\alpha_q = a_2 d_2 \frac{[IP_3]_i + d_1}{[IP_3]_i + d_3}, \quad \beta_q = a_2 [Ca^{2+}]_i. \qquad (4)$$

Where $[Ca^{2+}]$ represents the $Ca^{2+}$ concentration in the astrocyte cytoplasm. $[IP_3]$ is the concentration of the second messenger in astrocyte to modulate the opening dynamics of $Ca^{2+}$ channels on the astrocyte endoplasmic reticulum (ER). $q$ is the channel variable of the $Ca^{2+}$ channel, which is modulated by both the $[Ca^{2+}]$ and $[IP_3]$ in astrocyte cytoplasm. $[Ca^{2+}]_{ER}$ represents the equilibrium concentration of $Ca^{2+}$ in the ER. Other parameters that modulate the $Ca^{2+}$ channel dynamics in astrocyte have been explained in Table 1. In the Li-Rinzel model, the membrane leaky current $J_{leak}$ was introduced to represent the $Ca^{2+}$ ion efflux from astrocyte cytoplasm to the extracellular space. It has the mathematical form of $k_{leak}([Ca^{2+}]_i)$, where $k_{leak}$ represents the leaking rate and the negative symbol for $J_{leak}$ in Eq. (4) represents the efflux of $Ca^{2+}$ ions. $\xi_i(t)$ represents the zero mean, uncorrelated, and Gaussian white noise sources with $<\xi_i(t)\xi_i(t)>=2\alpha_q\beta_q/(\alpha_q+\beta_q)/N_q$, where $N_q$ is proportional to the number of ion channels in the ER membrane. Obviously, the channel noise intensity is proportional to $1/N_q$.

Glutamate-binding receptors located on the surface of the astrocyte membrane listen to abundant neuronal firing activities in response to sequential abrupt increases in IP3 and $Ca^{2+}$ in the cytoplasm. This process was modelled by Nadkarni et al.[20] with a dynamical equation to describe the proportional relation between the IP3 increase rate and released concentration of the neurotransmitter. However, to consider the heterogeneity among those G-protein coupled receptors in the astrocyte populations from a comprehensive perspective, we introduced noise into the model; thus, the final IP3 model has the following form:

$$\frac{d[IP3]_i}{dt} = \frac{([IP3]^* - [IP3]_i)}{\tau_{ip3}} + r_{ip3}\frac{([glu]_{neuron,i})^n}{([glu]_{neuron,i})^n + k^n} + DS * \xi_i^*(t) \qquad (5)$$
$$+ g_{astro-gap}(-(1+\tanh((|IP3_i - IP3_{i-1}|-0.3)/0.05))-(1+\tanh((|IP3_i - IP3_{i+1}|-0.3)/0.05)))$$

On the right of the IP3 dynamical model, the first term represents the decay process of intracellular IP3 due to enzymes, with the degradation time constant of $\tau_{ip3}$ and equilibrium concentration of $[IP3]^*$. The second term explains the increases in IP3 concentration modulated by glutamate binding to the G-protein receptors, with the binding intensity of $r_{ip3}$, and glutamate binding to the G-protein receptor follows the Hill formula to describe the saturation characteristics of the G-protein binding rate due to the increase in glutamate concentration. $k$ is the concentration at which IP3 production is halved[20]. For simplicity, the binding noise of neuronal glutamate onto the G-protein receptors, the IP3 changing noise, was defined as Gaussian white noise with an intensity of $DS$ to describe the heterogeneity between different astrocytes in a limited tissue in the network. The noise involves a standard distribution with the mean $<\xi^*(t)> = 0$, and the noise correlations are described by $<\xi^*(t)\xi^*(t')> = \delta(t-t')$.

The last term represents the gap-junctional connections among astrocytes are given by the nonlinear gap-junction model proposed by Goldberg et al.[42], which describes the threshold and nonlinearity of IP3 flow motivated by pulse-shaped $Ca^{2+}$ activities in astrocytes:

In the IP3 dynamical model, the integrated synaptic glutamate puff $[glu]_{neuron}$ represents the collective glutamate puff from all other neurons in the network, finally obtaining a summation form:

$$[glu]_{neuron,i} = (1-\lambda)\sum_{j=1,j\neq i}^{N} W_{ij}[T]_j(V_{s,j}), \qquad (6)$$

Where $\lambda$ represents the synaptic efficiency ratio of those glutamate that move to the postsynapses, and was set as 0.75[43], and 1-$\lambda$ corresponds to the partial ratio of those glutamates that successfully move to the astrocyte. $W_{ij}$ represents the connection state matrix value between the i$^{th}$ neuron and j$^{th}$ neuron: 1 means "connected state" and 0 means "not-connected state", and the connection probability is 0.22, which is close to the brain cortex parameter domain of 20% connection probability[44]. Because a large proportion of glutamate should get to the postsynapses. The glutamate puff from each neuron follows the law:

$$[T]_j(V_{s,j}) = 1/(1+exp(-(V_{s,j}-V_T)/\kappa)), \qquad (7)$$

To study the epileptic seizure of the network, which includes both the soma and dendrites, we utilized the two-compartment pyramidal models, including the dynamical membrane potentials of both the soma ($V_s$) and dendrite ($V_d$)[45]; see the Appendix for the detailed biophysical models of this Rinsky-Rinzel model. The two membrane potentials obtain their updated forms in the network, as shown in the following equations:

$$C_m \frac{dV_{s,i}}{dt} = I_{Leak}^s(V_{s,i}) - I_{Na}^s(V_{s,i},h_i) - I_{K-DR}(V_{s,i},n_i) + \frac{g_c}{p}(V_{d,i} - V_{s,i}) + \frac{I_{astro,i}}{p} \qquad (8)$$

$$C_m \frac{dV_{d,i}}{dt} = I^d_{Leak}(V_{d,i}) - I_{Ca_{neuron}}(V_{d,i}, s_i) - I_{K-AHP}(V_{d,i}, w_i) - I_{K-C}(V_{d,i}, c_i)$$
$$+ \frac{g_c}{1-p}(V_{s,i} - V_{d,i}) + \frac{I_{astro,i}}{1-p} + I_{syn,i} \qquad (9)$$

In the network model scheme, the coupling from each neuronal soma to the dendrite are coupled with an equivalent internal resistance $1/g_c*p$, and the internal resistance of the coupling from the dendrite to the soma is $1/g_c*(1-p)$. Each neuronal soma receives the input current from the feedback current of the neighbouring astrocyte ($I_{astro,i}/p$). Each ensemble dendrite receives feedback effects from the integrated synaptic current of the other network neurons ($I_{syn,i}$) and the feedback current of the neighbouring astrocyte ($I_{astro,i}/(1-p)$). The transmembrane ion channel currents in the soma and the dendrite can be described as a function of the corresponding potentials ($V_{s,i}$ and $V_{d,i}$) and the ion channel opening variables, whose mathematical equations are shown in the Appendix.

The astrocyte feedback current $I_{astro}$ in Eq. (8) involves the experiment-based model[19]:

$$I_{astro,i} = 2.11\Theta(\ln y)\ln y, \quad y = [Ca^{2+}]_i/nM - 196.69 \qquad (10)$$

The above model shows that this feedback current is highly dependent on the $Ca^{2+}$ dynamics of the neighboring astrocyte.

Meanwhile, the model scheme integrates synaptic inputs from other neurons in a homogeneous all−to−all random connection. The homogeneity is chosen for the consideration of weak distance-dependent effects on a small network scale of less than thousands of neurons. The homogeneous all−to−all random connection can be described by the integrated synaptic current:

$$I_{syn,i} = g_{syn}(\lambda \sum_{j=1, j\neq i}^{N} W_{ij}s_i)(V_{d,i} - E_{syn}), \qquad (11)$$

Where $W_{ij}$ represents the connection state between the i$^{th}$ neuron and j$^{th}$ neuron: 1 means "connected state" and 0 means "not-connected state" with the connection probability mentioned in Eq. (6). The connection map between neurons in one representative simulation trial is depicted in the upper right panel in Fig. 1.

The fraction of open channels *s* can be obtained by the dynamical equation of

$$\frac{ds_i}{dt} = \alpha_s \cdot \Theta(V_{s,i} - 20) - \beta_s \cdot s_i, \qquad (12)$$

where the Heaviside step function $\Theta(x)$ is known as the "unit step function", and its symbol can represent a piecewise constant function by the return whether *x* is over zero or not. The parameter $\alpha_s$ and rate $\beta_s$ represent values of the rise rate $\alpha_s$ and decay rate $\beta_s$ of the fraction s.

The biophysical definitions and values of all parameters in the above equations are listed in Table 1 in Appendix.

## III. THE TRANSITION ANALYSIS OF SEIZURE DYNAMICS

Generally, after the presynaptically released glutamate partially diffuses to the astrocyte that enwraps the postsynapse, the astrocyte responds to neuronal activation through an increase in the second massager IP3, but on the scale of neuronal networks, each astrocyte reacts differently even under the same neuronal stimulation due to the heterogeneity of the astrocyte G-protein receptors and extracellular circumstances. Here, in this paper, we utilized the noise intensity *DS* to represent the IP3 -reaction difference, detailed in the equations of the astrocyte IP3 dynamical models. The dynamic models in section 2 were simulated in the MALAB platform by using the Euler Method with the simulation timestep of 0.01ms.

The parameter *DS* was firstly discussed among 0, 0.25, 0.5, 0.75, 1.0, 1.25, 1.5, 1.75, and 2.0, during which we recorded all variables of neurons and astrocytes. A representative simulation trial of the action potential neuronal network was depicted in Fig. 2. It clearly presented evolutions of the spatiotemporal patterns of the neuronal network when *DS* increased.

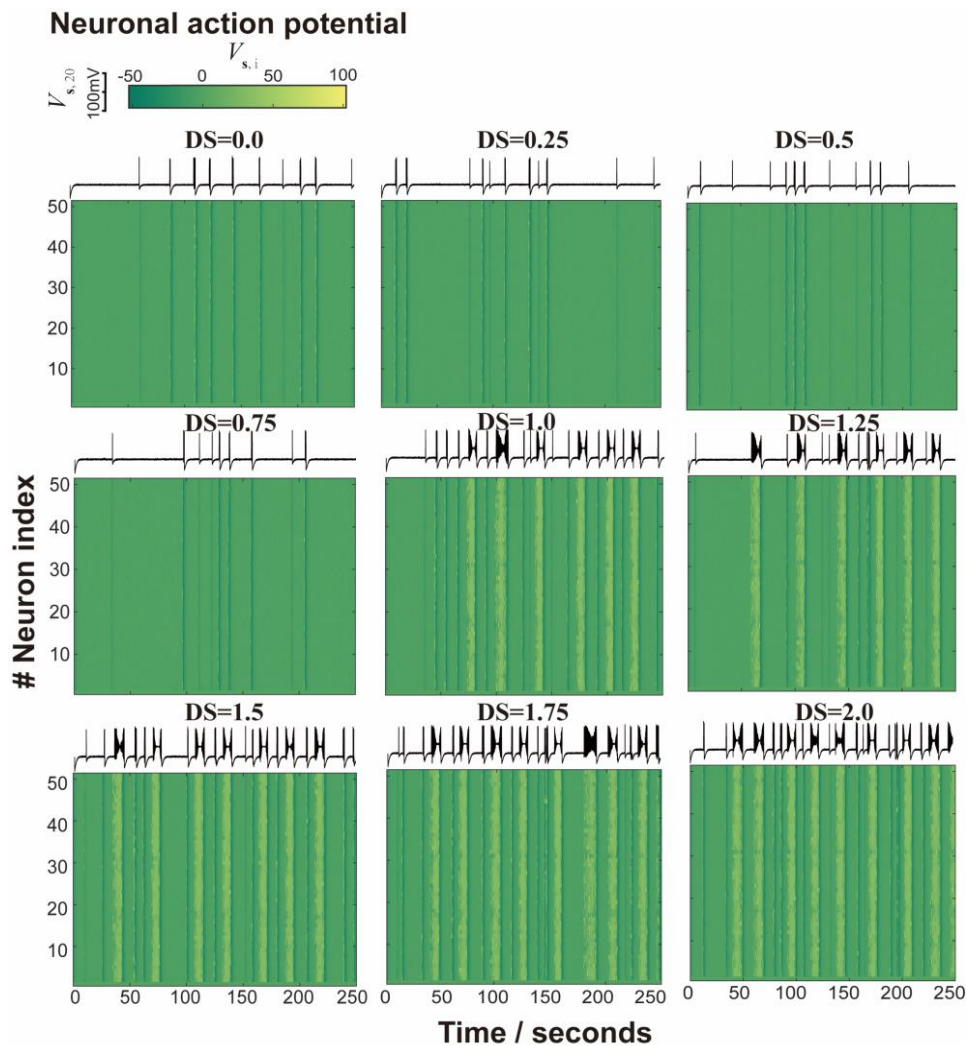

FIG. 2. Spatiotemporal firing patterns of the neuronal network exposed to astrocyte IP3 changing noise. The 9 subfigures in this figure correspond to the case where the noise intensity *DS* is equal to 0, 0.25, 0.5, 0.75, 1.0, 1.25, 1.5, 1.75, and 2.0. The single

neuronal firing in each subfigure is isolated from the 20th neuron in the network.

At first glance in Fig. 2, when *DS* was equal to 0~0.75, all neurons fired as a mixture of bursting and single spiking synchrony in the network. The bursting parts of firing tended to entail higher amplitude but randomly appeared in the network, as seen in the spiking pattern map when *DS* was equal to 0~0.75. In the resting state of the human brain, the firing of neuronal networks always oscillated relatively slowly(46). The neuronal network when *DS* was equal to 0~0.75 approximately reflected the focal firing pattern of the resting state. When *DS* jumped into the range of 1~2, the neurons started to fire in a seizure-like pattern, a dynamic switching between spiking and depolarization blocks (DBs). DB was defined as a typical seizure property with a glutamate-induced long-lasting depolarization phase and high rate of energy consumption[47]. Here, glutamate-induced DB was compromised with astrocytic gliotransmitters and neurotransmitters.

Meanwhile, 8 simulation trials of the single neuron firing patterns were presented when *DS*=0.25, 0.75, and 1.25 in Fig. 3 to discuss the random presence of epileptic seizures induced by the increase in IP3 changing noise. In Fig. 3, the red rectangles represented the epileptic depolarization blocks. As presented in Fig. 3 initially, in each case of *DS*, the neuron firing pattern always emerged differently across simulation trials but obtained a different distribution under different *DS* values. When *DS*=0.25, the neuron firing generally stayed in a sparse spiking pattern. This sparse firing pattern that was consistently absent with the depolarization blocks across different simulations could be the control state of the neuronal firing in comparison with the epileptic seizure state. Furthermore, when *DS* =0.75, the epileptic seizure started to be present in the neuronal network firing randomly across different simulations, in compared with the case when *DS* =0.25. Namely, in this condition of model parameters, both the epileptic seizure state and the sparse firing state had the potential to be induced. This could correspond to the brain state of the bi-stability when the *DS* started to affect the brain initially. Finally, when *DS* increased to 1.25, neuron firing persistently remained in the epileptic seizure state. This could reflect the fact that under this state of IP3 changing noise, the brain stayed in the epileptic state steadily.

Overall, Figs. 2 and 3 depicted clear trends for neurons transitioning from sparse spiking to epileptic-seizure events in three phases: a stable sparse spiking pattern, a bistable coexistence of sparse-spiking and epileptic-seizure patterns, and a stable epileptic-seizure pattern under the increase in astrocyte IP3 changing noise.
To accommodate the seizure results above, we analysed the collective firing pattern of the neuronal network in comparison with the experimental mean-field potentials of epileptic seizures[48]. Therefore, the mean voltage and the integrated synaptic current, together with single neuronal firing and astrocyte $Ca^{2+}$ and $K^+$ dynamics, were presented in Fig. 4. Because the dendrite branches were simplified into one dendritic model, the sum of the currents $I_{syn}$ represented the ensemble integrated mean field potentials of the focal neuronal network, namely, being capable of studying the mean-field potential characteristics. Meanwhile, the mean series of focal neuronal network firing $<V>$ represented the neuronal mean-field potentials. Fig. 4 showed the time series of single neuron (20[th]) firing, the ensemble synaptic current, and the mean voltages to be compared with the *in vitro* seizure in a single neuron and the extracellular mean field in Fig. 3 of Ref. 48. First, Fig. 4 showed that our simulated seizure in a single neuron was composed of sparse spiking and epileptic depolarization blocks that generally persisted at a stable value for a short time, which was close to the *in vitro* seizure firing of single neurons[48]. In addition, in the collective view of the neuronal network dynamics, the simulated mean voltages closely resembled the time series of the ensemble synaptic current in Fig. 4, and they both involved a potential peak at the time towards the single neuronal depolarization block phase. Surprisingly, this correspondence between a single neuron and the mean field could also be found in *in vitro* seizures in Fig. 3(a) of Ref. 48. On the one hand, this vigorously showed the representative capacity of the synaptic current and the mean voltages for focal mean-field potential; on the other hand, the consistent presence of the correspondence between a single neuron and the mean field in both the simulation and *in vitro* results indicated that the results in this paper were provable.

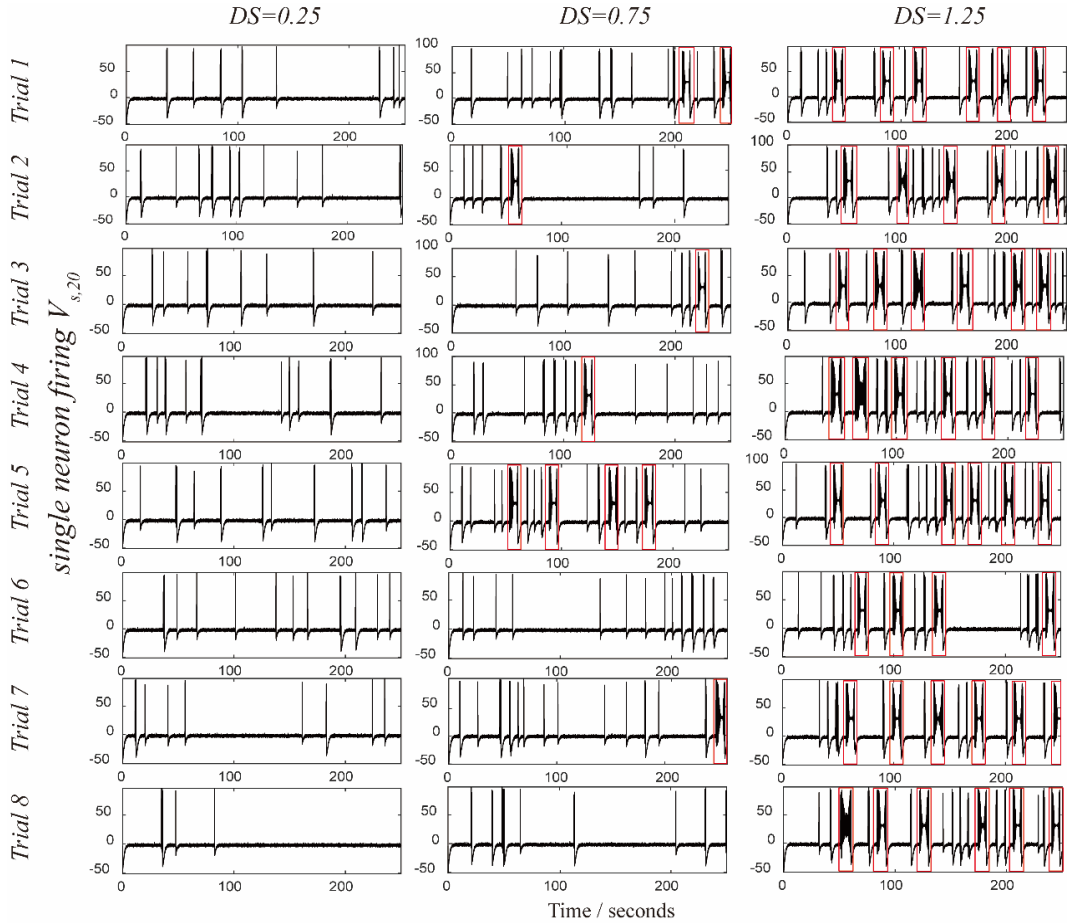

FIG. 3. The coexistence phenomena of the epileptic DB and the regular spiking under the critical parameter of inositol triphosphate noise intensity *DS*. The total number of simulation trials for each condition of *DS* was 60, and only 8 random trials of single neuronal firing $V_{s,20}$ for each *DS*= 0.25, 0.75, and 1.25 were present for comparison. All epileptic specific depolarization block phases during seizures are circled with red rectangles.

Furthermore, the intracellular astrocytic $Ca^{2+}$ and extracellular $K^+$ were also studied in contrast with epileptic seizure potentials in this paper. The results were shown at the bottom of Fig. 4. First, it was very easy to determine that the depolarization block phase in the seizure potentials parallels the peaks of $Ca^{2+}$ and $K^+$, which reflected the experimental results indicating that depolarization-block seizure firing induced much more glutamate released onto astrocytes and sequentially excited astrocytes with much higher IP3 and $Ca^{2+}$ concentrations. Meanwhile, the depolarization block phase of seizure potentials gave rise to inward $Na^+$ channels but inhibited outward $K^+$ channels. This directly rendered extracellular spaces with a much higher $K^+$ concentration. In light of the above modulations of astrocyte $Ca^{2+}$ and $K^+$ in neuronal firing, astrocytes were a very sensitive partner to sense and responded to neuronal seizure potentials in terms of $Ca^{2+}$ and $K^+$.

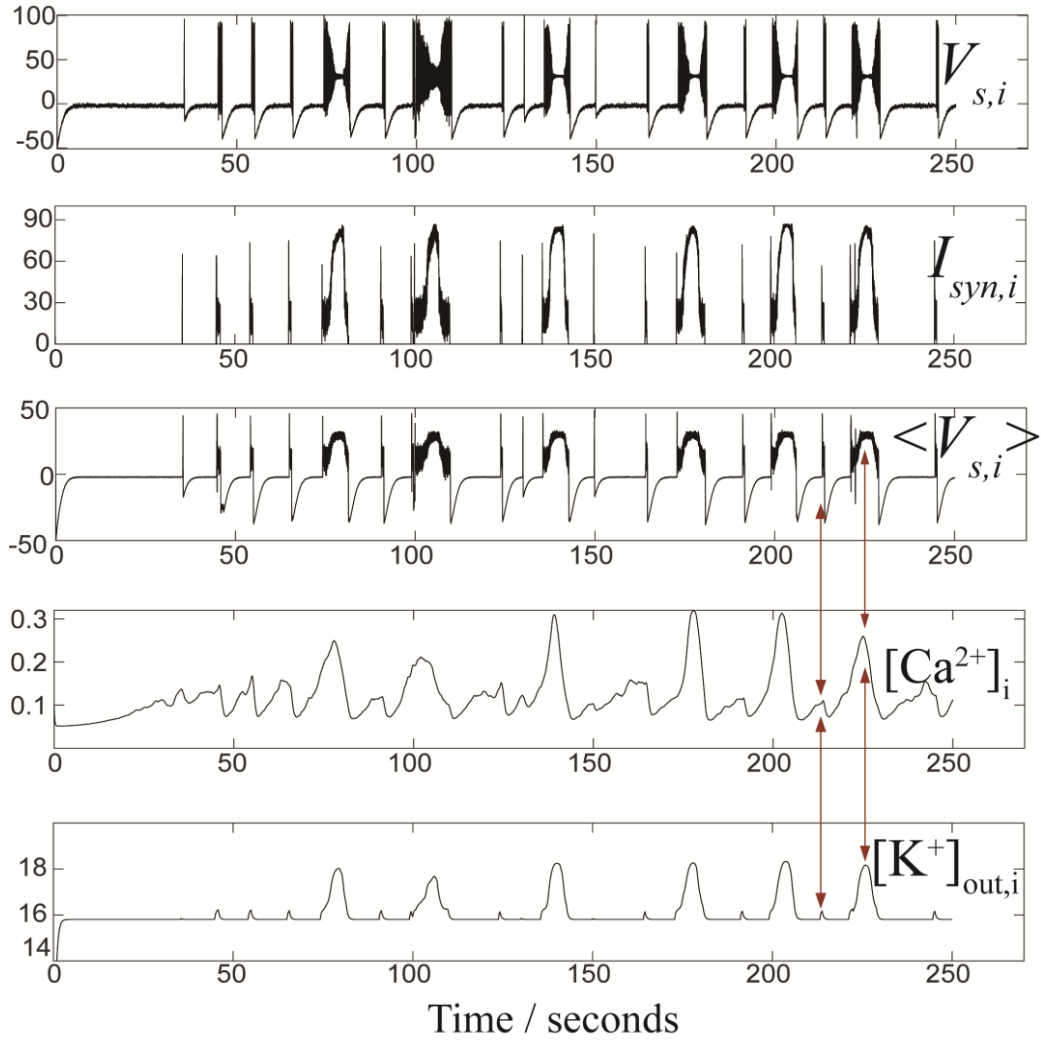

FIG. 4. The slow modulation of astrocyte oscillation pattern $Ca^{2+}$ and $K^+$ dynamics in the assembly fast synaptic current $I_{syn}$ and mean voltage $<V>$ during the epileptic seizure condition with $DS=1.0$.

Seizures generally involve dynamical specificities not only in amplitude, such as the results shown in Fig. 4, but also in firing frequency. Therefore, we studied the ensemble spectral power distribution during the frequency domain 0~500 Hz and the time-frequency analysis based on the *N* neuronal potentials in cases when the noise intensity *DS*=0 and *DS*=1.0. The simulation results were presented in Fig. 5. First, to detect the high-frequency oscillation (HFO) characteristics that were found to distribute among 80~500 Hz[49], the frequency band of 0~ 500 Hz was introduced to analyse the down-sampled 1000 Hz neuronal voltage data by the Fourier transformation method. Fig. 5(a) first showed the spectral power distributional difference across frequencies when *DS*=0 and *DS*=1.0. We could easily determine that the voltage data with IP3 noise interpretation exhibited much higher spectral power than the no-noise case over the frequency range of approximately 80~400 Hz. This gave a first glance at seizure events induced by IP3 changing noise that involved the epileptic experimental high-frequency oscillation and HFO characteristics. Moreover, the dynamic evolution of the voltage frequency spectral power was studied to investigate the original components of those HFOs, as shown in Fig. 5(a). First, Fig. 5(b) showed that when *DS*=1.0, there were concentrated high-frequency timings in addition to the moments when the seizure depolarization block occurred. However, when *DS*=0, these concentrated high-frequency timings were not present during the entire time range. This comparably implied that seizure depolarization could be a "HFO attractor" that finally induced specific HFO phenomena in epileptic seizures.

Generally, epileptic seizures occur randomly; however, it has been recently demonstrated that epileptic patients convey bi-stable states of both regular and epileptic seizures, and external stimulation could induce the transition between the two states[50]. Therefore, epileptic seizures conferred dynamic stability characteristics, which corresponded to the epileptic depolarization block in this paper. To measure the stable regulations of the epileptic DBs and determine the critical state of the epileptic seizure in the stochastic process, we introduced the seizure DB information entropy. The formula of the entropy computation method was placed in the Appendix to statistically represent the seizure induction state and utilize the t-test among 60 simulation trials in the *DS* condition to determine the significance of each case of *DS* compared with the control case *DS*=0. Entropy theory was generally used to describe the probability distributions of spiking patterns (words) in a series of simulated or experimental potentials[51–54], and it was more efficient to distinguish ignorable changes in neural activities by analysing neuronal information entropy theory than neuronal firing information directly[55]. In this paper, the intended "words" were the DBs, and the entropy of this potential series represented the seizure-inducing degree in one case of IP3 changing noise. Meanwhile, the final DB entropy was obtained by averaging all entropy values across all neuron cells. The simulation results of the DB entropy versus IP3 changing noise were depicted in Fig. 6.

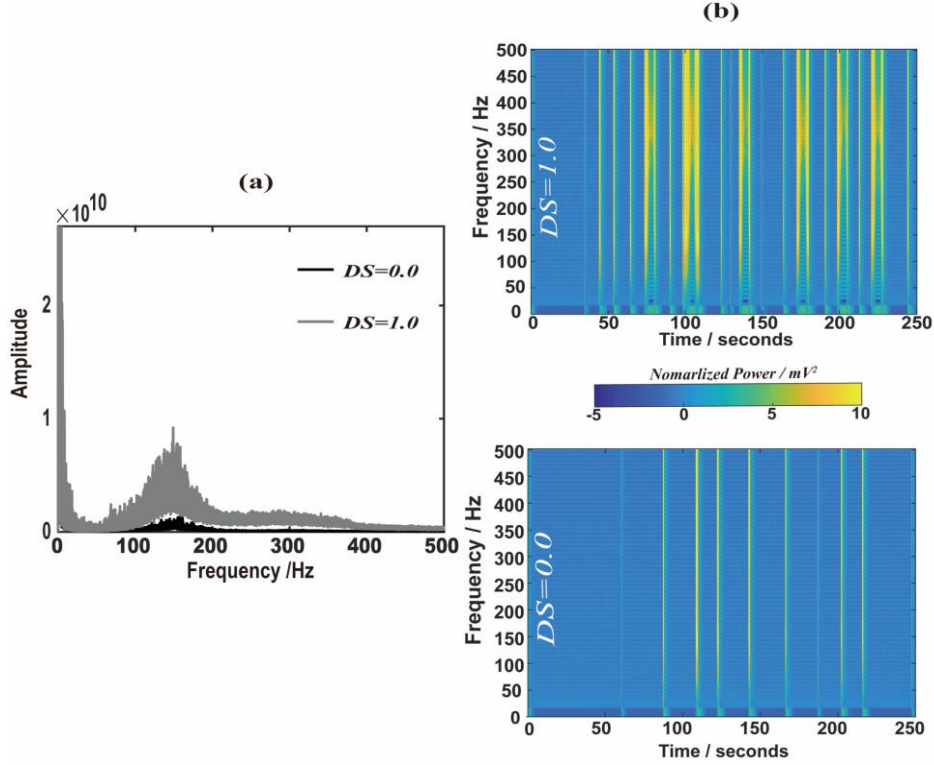

FIG. 5. Noise-induced neuronal frequency switches between low and HFO bands. (a) The spectral power of the neuronal network firing averaged across different neurons when *DS*=0.0 and 1.0; (b) the normalized frequency versus time map when *DS*=0.0 and 1.0. The normalized power values were obtained using a fast Fourier transform with a nonoverlapping window of 0.125 s (the total time was 250 s) and then averaged across neurons and normalized versus the baseline data from 0~10 s.

At first glance, we found in Fig. 6 that the DB entropy gradually increased when *DS*≥0.75 and finally stabilized at approximately *DS* =1.25. This implied that when the *DS* increased initially, epileptic seizure events started to be induced but stayed in a state with fewer DBs; however, when the noise intensity gradually grew, epileptic seizure events induced more frequent DBs. Meanwhile, statistical analysis provided us with information showing that epileptic seizure events were significantly induced when *DS* ≥0.75. However, we also found that epileptic seizure events were absent in some simulation trials when *DS*=0.75, which exactly predicted this state of criticality for the epileptic system of neurons and astrocytes. The DB presence at each *DS* value could be obtained based on the DB spatiotemporal patterns in Fig. 6.

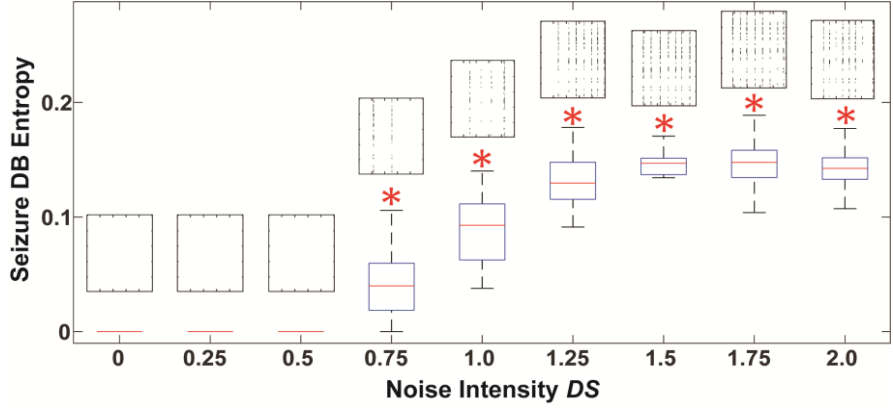

FIG. 6. The seizure DB map and entropy versus the noise intensity $DS$. The DB entropy is obtained based on the DB map data by using the information equation in the Appendix. The zoomed box map for each $DS$ state represents the DB map, where $x$ axis is the time ranging from 0 to 250 seconds, and $y$ axis is the neuron index ranging from 0 to 50, similar with voltage maps in Fig. 2. The seizure DB map is obtained by filtering other voltage signals by defining a "1" for a sliding of 1 s as the difference between the maximal and minimal voltage values from this time window and "0" for other cases. Because of the randomness of seizure induction, 20 trials for each case of DS were performed, and a t-test was applied to all trials of $DS$=0.25, 0.5, 0.75, 1.0, 1.25, 1.5, 1.75, and 2.0 versus the control case $DS$=0 with Bonferroni correction =0.025/8. The "*" indicates that the DB entropy of one $DS$ case versus the control is statistically significant.

Furthermore, for a biological investigation of the stochastic process of epileptic seizures under IP3 changing noise, we measured the DB durations in cases with different noise intensities. The DB duration statistics in Fig. 7 were performed by averaging all DBs across all neuron cells and depicting the mean value and each point of the DBs from all simulation trials at each IP3 changing noise condition. All DBs computed in seconds were consistent with related experiments determining seizure duration measurements[56].

We could see from Fig. 7 that the DB duration stayed steadily zero when $DS$<0.75; however, when $DS$ continued to increase, the DB duration elevated into a region from approximately 10 to 30 seconds. Based on a previous experimental study, the simulation DB duration satisfied the seizure-significant presence region (from approximately 10 to 60 seconds) in the experiment[56]. In summary, Fig. 6 showed the formation process of a stable epileptic seizure state using the Shannon entropy method, while Fig. 7 showed the statistical analysis of the degree to which stable epileptic seizures occurred during DB durations. In addition, this could partly indicate that our simulation method was trustworthy for elucidating the stochastic process of the epileptic seizure induction process.

As demonstrated in earlier experimental studies, astrocytes responded to both epileptic seizures and external cognitive tasks with long-lasting $Ca^{2+}$ signals in intracellular astrocytes[57–63]. However, there remained some confusion regarding the dynamic characteristics after certain conditional stimuli, as the previous studies found abundant temporal patterns of $Ca^{2+}$ signals in astrocytes in different temporal scales etc.[64, 65]. Here, to unfold the critical role of IP3 receptor-dependent astrocyte $Ca^{2+}$ signal in the neuronal behavior of epileptic seizures, we studied the $Ca^{2+}$ signal changes in frequency and amplitude under the condition of IP3 changing noise when $r_{ip3}=0$ (IP3-/-) and $r_{ip3}=0.28$ (IP3 nonblock). The frequency coding was extracted by performing Fourier transformation for the simulation data of all trials and taking the mean of the frequency power in Fig. 8(a) and then depicting the mean and standard deviation of the maximal frequency of each simulation trial together with the redrawn experimental data[63] in Fig. 8(b). The spatiotemporal patterns of the $Ca^{2+}$ signals in the case of both IP3-/- and IP3 nonblock mice and the phase map of astrocytic $Ca^{2+}$ systems were depicted in Fig. 8(c), d to explain the multiple-amplitude-state phenomena of $Ca^{2+}$ signals under epileptic seizures.

First, Fig. 8(a) showed that the frequency of the $Ca^{2+}$ signals tended to concentrate to approximately 0.05 Hz when the G-protein-bound receptor was not blocked (IP3 nonblock) but concentrated to approximately 0.01 Hz when the IP3 receptor was blocked. We counted the mean and standard deviations of the peak frequency of Fig. 8(a) and presented them in unit of /min, in contrast with the corresponding case of IP3-/- and IP3 nonblock based on the experimental data[63]. From Fig. 8(b), we could determine that in the case of IP3 nonblock, the maximal difference between the means of the experimental and simulation data were 1 $min^{-1}$. This minor mismatched error could arise from other uncovered models of non-IP3 pathway-induced $Ca^{2+}$ signals, such as voltage-gated calcium channels (VGCCs)[66, 67] and store-operated calcium entry[68] in vivo. In conclusion, the results above indirectly implied that our established models in this paper were very helpful and reproducible for studying the activities of epileptic seizures and astrocytes in vivo.

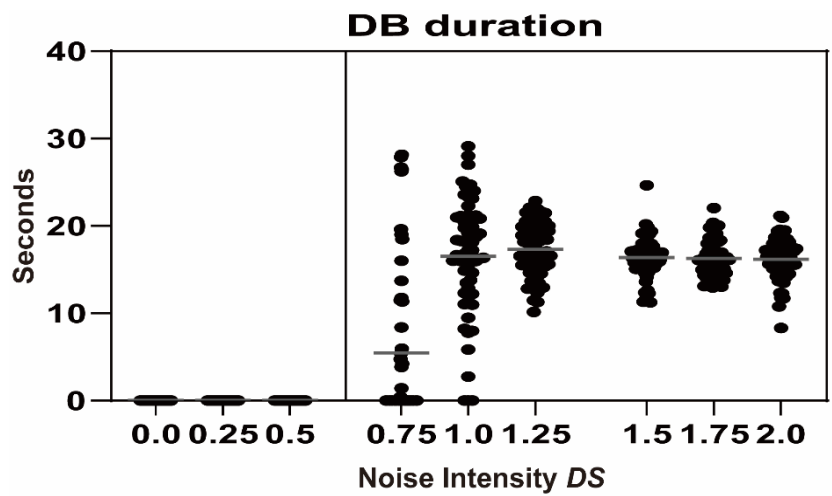

FIG. 7. Noise-induced seizure DB duration bifurcation versus the noise intensity. The DB duration of all 60 trials of each *DS* is depicted in the figure, and each DB duration value in the figure was obtained by computing the interval between two DB starts and then calculating the average across the neuron cells. The bistable states of seizure dynamics are present in the conditions when *DS*=0.75, and 1.0.

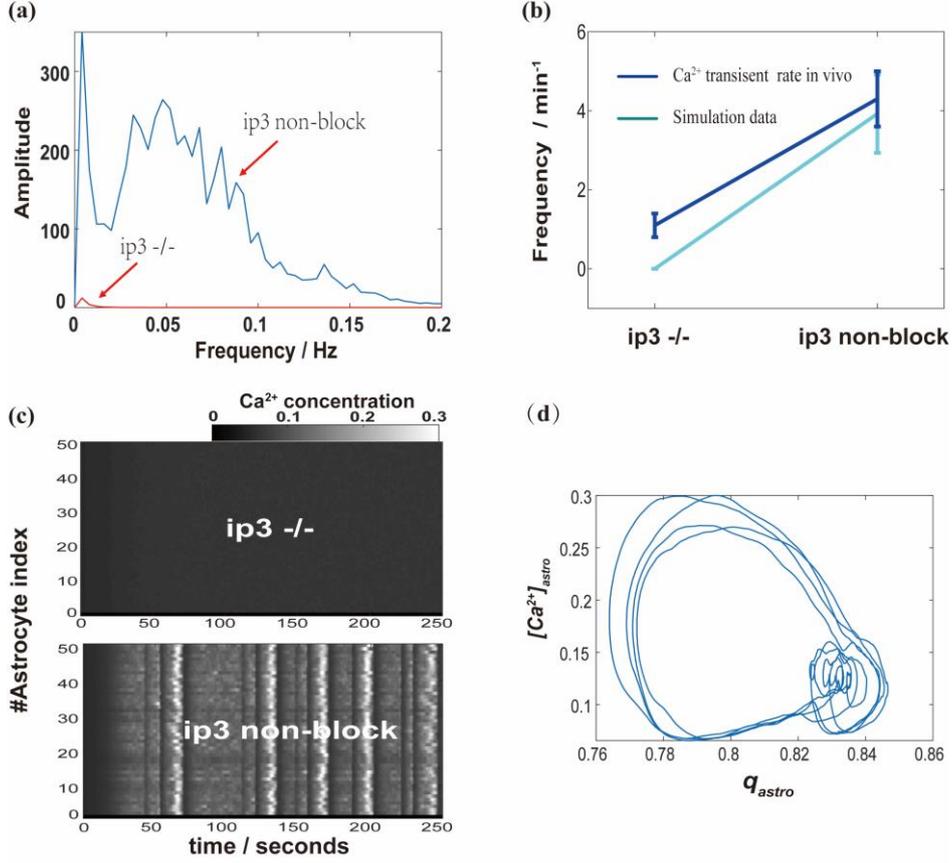

FIG. 8. The slow dynamics of the astrocyte $Ca^{2+}$ transient in its frequency domain in response to the inositol triphosphate noise. (a) The simulation results of the spectral power of $Ca^{2+}$ transients in the astrocyte network when $r_{ip3}$=0 (IP3-/-) and $r_{ip3}$=0.28 (IP3 nonblock); (b) the mean rate of $Ca^{2+}$ transients in the astrocyte network when IP3 stays blocked and nonblocked in the conditions of in vivo versus simulation results, respectively. The in vivo data were obtained by analyzing the histograms of the Fig. 3(c) left panel from the Ref. 63 with the Origin Digitizer; (c) the population patterns of the $Ca^{2+}$ transients in the astrocyte network when $r_{ip3}$=0 (IP3-/-) and $r_{ip3}$=0.28 (IP3 nonblock); (d) the phase map of astrocyte $Ca^{2+}$ signals and the gating variable $q_{astro}$ when $r_{ip3}$=0.28 (IP3 nonblock).

## IV. CONCLUSIONS AND DISCUSSIONS

The randomness of epileptic seizures has been a significant obstacle for the prediction and treatment of epilepsy. To investigate the stochastic process of seizure generation, we constructed tripartite synapse network models of neurons and astrocytes that exhibit IP3 changing noise. Our simulation results have shown that the dynamics of epileptic seizures start to emerge significantly when the IP3 changing noise intensity *DS* increases over a critical state. Surprisingly, at the critical value of *DS*, the neurons obtain concomitant patterns of both sparse firing and epileptic seizures. Generally, the severe abnormality of brain damage after seizures always comes with a higher rate of epileptic seizures for certain subjects[69], the above results under the increase of *DS* indirectly reflects that the brain is suffering from potential brain damage to affect the noise discussed in this paper. In addition, the positive role of astrocytes in epileptic seizures has been shown in this paper. We proved that astrocytic $Ca^{2+}$ signals tend to be activated after the increase in *DS*, accompanied by a peak $K^+$ concentration. Additionally, $Ca^{2+}$ and $K^+$ can sustain the generation of depolarization blocks in a single neuron and the abrupt mean field pulse at the time when $Ca^{2+}$ and $K^+$ peak. To investigate the duration changes of the epileptic seizure under IP3 changing noise and verify the reliability of epileptic seizure phenomena shown in this paper, the statistical DB duration among all simulation trials was studied versus the noise intensity *DS*, in contrast with the duration measurements in the experiment[56]. The results have shown that as the *DS* increased over a threshold, the epileptic seizure started to be induced, and then larger *DS* would induce a more stable distribution of DB durations. Considering the $Ca^{2+}$ signal frequency, we further verified the IP3-astrocyte $Ca^{2+}$ signals that sustain the epileptic seizure in this paper in a reliable parameter region by comparing the $Ca^{2+}$ signal frequency of IP3-/- and IP3 nonblock mice between the simulation and experimental cases. The results support our speculation that the $Ca^{2+}$ signal frequency is located in a reliable region, as observed in our experiments, further suggesting the reliability of our proposed network model for describing neuronal firing characteristics in vivo.

Virtually all of the results above in this paper simulated the stochastic process of epileptic seizure induction under IP3 changing noise. The simulation model and the results would be potential references for controlling the epileptic seizures by modulating the noise-related factors.

However, some points related to the dynamics of epileptic seizures are still worth investigating in future studies. For example, because of the importance of GABAergic interneurons in the "shutdown" of epilepsy, studies of GABAergic interneurons have improved them to one of the top positions in anti-epileptic programs[70–72]. Therefore, it is necessary to unveil the underpinnings of GABAergic interneurons in inhibiting the spread of epileptic seizures by modulating astrocytes in simulation methods. In addition, the energy consumption of epileptic seizures in human epileptic subjects was proven to be distinguishable when compared with the resting state using the combined methods of fMRI and EEG[73]; therefore, unveiling the dynamic mechanism underlying how astrocytes transfer glucose into neuronal circumstances is significant for achieving a better understanding of the energy consumption theory of epileptic neuronal systems. Finally, it is known that febrile convulsions are generally induced with epileptic seizures by increasing brain temperature[74, 75]. This could be one of the factors that affect the random binding dynamics of neuronal glutamate to G protein-bound receptors; thus, it is worth utilizing a stochastic molecular dynamics simulation algorithm, e.g., the Monte Carlo method, to study the stochastic mechanism of febrile convulsion.

**Appendix: The two-compartment neuron model and basic model parameters**

The conductance-based Pinsky−Rinzel model[44] is described by the following set of ordinary differential equations:

$$C_m \frac{dV_s}{dt} = I^s_{Leak}(V_s) - I^s_{Na}(V_s,h) - I_{K-DR}(V_s,n) + \frac{g_c}{p}(V_d - V_s) \quad (A1)$$

$$C_m \frac{dV_d}{dt} = I^d_{Leak}(V_d) - I_{Ca_{neuron}}(V_d,s) - I_{K-AHP}(V_d,w) - I_{K-C}(V_d,c) + \frac{g_c}{1-p}(V_s - V_d) \quad (A2)$$

where the transmembrane ion channel currents in the soma and the dendrite can be described as a function of the corresponding potentials ($V_s$ and $V_d$) and the ion channel opening variables, whose mathematical equations are shown below:

The somatic ion currents:

$$I_{Na}(V_s,h) = g_{Na} m_\infty^2 h (V_s - V_{Na})$$

$$I_{K-DR}(V_s,n) = g_{K-DR} n (V_s - V_K) \quad (A3)$$

$$I^d_{Leak}(V_d) = g_L(V_d - V_L)$$

The dendrite ion currents:

$$I_{Ca_{neuron}}(V_d,s) = g_{Ca_{neuron}} s^2 (V_d - V_{Ca_{neuron}})$$

$$I_{K-AHP}(V_d,w) = g_{K-AHP} w (V_d - V_K) \quad (A4)$$

$$I_{K-C}(V_d,c) = g_{K-C} c \chi([Ca_{neuron}])(V_d - V_K)$$

$$\chi([Ca_{neuron}]) = \min(\frac{[Ca_{neuron}]}{250}, 1.0)$$

In the above equations, the gating variables of $h$, $n$, $s$, $w$ adapt to the biophysical models of

$$\frac{dy}{dt} = \alpha_y(x)(1-y) - \beta_y(x)y \quad (A5)$$

$$x = \begin{cases} V_s & y = h,n,m \\ V_d & y = s,c \\ Ca_{neuron} & y = w \end{cases}$$

The rise rate $\alpha_y$ and decay rate $\beta_y$ of the gating variables are obtained by experimental adaptions in the mathematical forms of

$$\alpha_m = \frac{0.32(13.31-V_s)}{\exp((13.31-V_s)/4)-1}, \beta_m = \frac{0.28(V_s - 40.1)}{\exp((V_s - 40.1)/5)-1}$$

$$\alpha_n = \frac{0.016(35.1-V_s)}{\exp((35.1-V_s)/5)-1}, \beta_n = 0.25\exp(0.5 - 0.025V_s)$$

$$\alpha_h = 0.128\exp((17-V_s)/18), \beta_h = \frac{4}{1+\exp((40-V_s)/5)}$$

$$\alpha_s = \frac{1.6}{1+\exp(-0.072(V_d - 65))}, \beta_h = \frac{0.02(V_d - 51.1)}{\exp((V_d - 51.1)/5)-1}$$

$$\alpha_c = \begin{cases} (\exp((V_d - 10)/11 - (V_d - 6.5)/27))/18.975 \\ 2\exp((V_d - 6.5)/27) \end{cases}$$

$$\beta_c = \begin{cases} 2\exp((V_d - 6.5)/27) - \alpha_c/18.975 & V_d \leq 50 \\ 2\exp((V_d - 6.5)/27) & V_d > 50 \end{cases}$$

$$\alpha_w = \min(0.00002[Ca_{neuron}], 0.01), \beta_w = 0.001$$

where $[Ca_{neuron}]$ is the calcium concentration at the dendrite and modulated by the dendrite potential. The mathematical form of $[Ca_{neuron}]$ can be described by the following equation:

$$\frac{d[Ca_{neuron}]}{dt} = -0.13 g_{Ca} s^2 (V_d - V_{Ca}) - 0.07[Ca_{neuron}] \quad (A6)$$

In this paper, we developed the potassium reversal potential $V_K$ as a variable instead of a constant value based on the method of Fröhlich [50]. The kinetic dynamical equations are described by the Nernst equations and involve the mathematical forms of

$$V_K = 26.64 \, \log((K_{out}/K_{in})) \quad (A7)$$

Here, $K_{in}$ was defined as a constant value of 130 mV for both somas and dendrites, $K_{out}$ represents the accumulation of both soma and dendrite potassium channel efflux and has the dynamical equation:

$$\tau_K \frac{dK_{out,i}}{dt} = \kappa_K I_{K_{efflux,i}} - (J_{lateral-duffusion,i} + J_{bath-diffusion,i}) - J_{pump,i} - J_{astro,i} \quad (A8)$$

In the above equation left, $\tau_K$ represents the biophysical time constant of the potassium concentration in the extracellular space.
The first term in the equation on the right presents the total potassium efflux from the soma and dendrite and involves the mathematical forms of $(I_{K-DR} + I_{K-C} + I_{K-AHP})$. The multiplication parameter $\kappa_K = zFV_{ast}$ denotes the physical scaling constant between the current and the potassium ion flux[66].

$J_{pump,i}$ arises from a backwards influx from the extracellular space (ECS) and negatively contributes to $K_{out,i}$. It has the mathematical form of $35/(1.0+(3.5/K_{out,i}))^{2}$ [50]. Meanwhile, astrocytes in the local circuit environment still take up potassium via Kir4.1 and water channels, and this process of $J_{astro,i}$ was modelled as $g_{astro\_max}/(1.0+\exp((18.0-K_{out,i})/2.5))$ [50].

A local potassium ion can diffuse in two ways: between different local circuits and between the local cell circuit and the base potassium environment outside the current network of neurons and astrocytes; therefore, the intercellular lateral diffusion flux $J_{lateral-ffusion,i}$ and the base-environment dependent potassium flux $J_{base-diffusion,i}$ have been added into the $K_{out}$ dynamical model[17], and they obtained the mathematical forms of $\varepsilon_K (K_{out,i-1} - 2K_{out,i} + K_{out,i+1})$ and $\varepsilon_K (K_{out,i} - K_{bath})$, respectively;

**The entropy theorem:**
In this paper, the Shannon entropy equation was introduced to provide a method to estimate the average minimum number of bits needed to encode a string of epileptic depolarization blocks, or DB symbols, based on the frequency of the DB symbols. As shown in the DB maps of the inserted figures in Fig. 6 in the manuscript, all neuronal firing in one noise intensity *DS* case was transformed into 1-0 signals: for each nonoverlapping time window of neuronal firing, if the window voltage signal remained steady with a small noisy amplitude of 5 mV, we defined the transformed signal as 1 at this time window; otherwise, it was 0 based on the *in vitro* epileptic depolarization block characteristics[48]. For each transformed neuronal firing time series, the epileptic

DB entropy can be described by the equation:

$$H_{DB} = -\sum_{i=1}^{N-1} p_i \log_2 p_i \qquad (A9)$$

where $p_i$ is the presence probability of a given DB symbol in each transformed series of neuronal firing. Finally, the DB entropy in Fig. 6 was obtained by averaging all $H_{DB}$ values across all neurons in the network.

**TABLE 1.** Parameter values for the tripartite synaptic neuron-astrocyte network model and the Pinsky–Rinzel model.

| Parameter | Value | Parameter | Value |
|---|---|---|---|
| $C_m$ | 3µFcm$^{-2}$ [44] | $d_1$ | 0.13µM [41] |
| $g_{Na}$ | 30 mScm$^{-2}$ [44] | $d_2$ | 1.049 µM [41] |
| $g_{K-DR}$ | 15 mScm$^{-2}$ [44] | $d_3$ | 0.9434µM [41] |
| $g_{k-AHP}$ | 0.8 mScm$^{-2}$ [44] | $d_5$ | 0.08234µM [41] |
| $g_{K-C}$ | 15 mScm$^{-2}$ [44] | $k$ | 0.785 [20] |
| $g_{NaL}$ | 0.05mScm$^{-2}$ [44] | $N_{glu}$ | 0.45 [20] |
| $g_L$ | 0.05 mScm$^{-2}$ [44] | $g_{astro-gap}$ | 2µM s$^{-1}$ [42] |
| $g_{Ca}$ | 10 mScm$^{-2}$ [44] | $V_T$ | 5mV [44] |
| $g_C$ | 2.1mScm$^{-2}$ [44] | $K$ | 2s [44] |
| $V_{Na}$ | 120mV [44] | $g_{syn}$ | 0.045mS cm$^{-2}$ [44] |
| $V_K$ | -15mV [44] | $E_{syn}$ | 140mV [44] |
| $V_L$ | 0mV [44] | $\alpha_s$ | 1s$^{-1}$ [44] |
| $V_{Ca}$ | 140mV [44] | $\beta_s$ | 0.5 s$^{-1}$ [44] |
| $P$ | 0.5 [44] | $Z$ | 2 [66] |
| IP3* | 0.16µM [19] | $F$ | 96 485 C/mol [66] |
| $\tau_{IP3}$ | 7.14s [19] | $V_{ast}$ | 5.233×10$^{13}$ l [66] |
| $r_{IP3}$ | 0.8µM s$^{-1}$ [19] | $g_{astro\_max}$ | 66 mM s$^{-1}$ [50] |
| $c_1$ | 0.185 [41] | $\varepsilon_k$ | 1.2 s$^{-1}$ [50] |
| $v_1$ | 6 s$^{-1}$ [41] | $K_{kbath}$ | 4.0 mM [50] |
| $v_2$ | 0.11 s$^{-1}$ [41] | $K_{in}$ | 130mM [50] |
| $v_3$ | 0.9µM.s$^{-1}$ [41] | $N_q$ | 50 [76] |
| $a_2$ | 0.2µM.s$^{-1}$ [41] | | |

**AUTHOR'S CONTRIBUTIONS**

**Jiajia Li**: Conceptualization (lead); writing – original draft (lead); formal analysis (lead); Methodology (equal); writing – review and editing (equal). **Peihua Feng**: Methodology (equal); writing – review and editing (equal). **Liang Zhao**: Software (equal); writing – review and editing (equal). **Junying Chen**: Software (equal); writing

– review and editing (equal). **Mengmeng Du**: Writing –review and editing (equal). **Jian Song**: Conceptualization (supporting); Writing – original draft (supporting); Writing – review and editing (equal). **Ying Wu**: Writing –review and editing (equal).

ACKNOWLEDGEMENT

We would like to thank all participants who were enrolled in our study. This work was supported by the National Natural Science Foundation of China (Grant Nos. 12002251, 1197020421, 12102240), the Natural Science Foundation of Shaanxi Province (Grant Nos. 2020JM-473) and the Xi'an Science and Technology Program (Grant No. 2020KJRC0055), Scientific Research Program Funded by Shaanxi Provincial Education Department (Program No. 21JK0710).